\documentclass[amsmath,showpacs,pre,reprint,superscriptaddress,longbibliography]{revtex4-1}
\usepackage{txfonts}
\usepackage{graphicx}
\usepackage[colorlinks]{hyperref}

\begin{document}
\title{Clique percolation in random graphs}

\author{Ming Li}
\affiliation{Department of Modern Physics, University of Science and Technology of China, Hefei, 230026, People's Republic of China}

\author{Youjin Deng}
\affiliation{Hefei National Laboratory for Physical Sciences at Microscale, University of Science and Technology of China, Hefei, 230026, People's Republic of China}
\affiliation{Department of Modern Physics, University of Science and Technology of China, Hefei, 230026, People's Republic of China}

\author{Bing-Hong Wang}
\affiliation{Department of Modern Physics, University of Science and Technology of China, Hefei, 230026, People's Republic of China}

\date{\today}

\begin{abstract}
As a generation of the classical percolation, clique percolation focuses on the connection of cliques in a graph, where the connection of two $k$-cliques means that they share at least $l<k$ vertices. In this paper, we develop a theoretical approach to study clique percolation in Erd\H{o}s-R\'{e}nyi graphs, which gives not only the exact solutions of the critical point, but also the corresponding order parameter. Based on this, we prove theoretically that the fraction $\psi$ of cliques in the giant clique cluster always makes a continuous phase transition as the classical percolation. However, the fraction $\phi$ of vertices in the giant clique cluster for $l>1$ makes a step-function-like discontinuous phase transition in the thermodynamic limit and a continuous phase transition for $l=1$. More interesting, our analysis shows that at the critical point, the order parameter $\phi_c$ for $l>1$ is neither $0$ nor $1$, but a constant depending on $k$ and $l$. All these theoretical findings are in agreement with the simulation results, which give theoretical support and clarification for previous simulation studies of clique percolation.
\end{abstract}


\maketitle

\section{Introduction}

Percolation theory has been widely used in the study of complex systems and complex networks\cite{newman2010networks}. In general, percolation theory considers the emergence of the giant cluster formed by occupied vertices or edges in a graph, when possible vertices or edges are occupied with a given probability $0<p<1$. The characteristics of percolation transitions in different graphs can give many useful results for the structural design and dynamic analysis of real network systems. For example, the percolation transition on graphs with degree distribution $p_k\sim k^\gamma$ ($2<\gamma<3$) shows that such a system has a critical point $p_c\rightarrow 0$ for random vertex occupation and a very large $p_c$ for targeted vertex occupation (unoccupied large degree vertices with a high probability)\cite{PhysRevLett.85.4626,PhysRevLett.85.5468,PhysRevLett.86.3682}. This indicates that real network systems with such a degree distribution are robustness against random failure, but extremely fragile against a targeted attack.

In addition, some modified percolation models have been introduced into the study of complex networks, such as $k$-core percolation\cite{PhysRevLett.96.040601}, bootstrap percolation\cite{Watts30042002}, and percolation in independent networks\cite{Buldyrev2010}. All these models play an important role in the theoretical studies of the cascade and diffuse processes on networks. At the same time, some new phenomena have also been found in the modified percolation models\cite{EPJ2014,Saberi20151}, such as the discontinuous percolation transition\cite{PhysRevLett.96.040601,Watts30042002,Buldyrev2010}.

To identify overlapping communities in large real networks, Der\'{e}nyi and co-workers have proposed a dependent percolation model, called clique percolation\cite{palla2005uncovering,PhysRevLett.94.160202}. A clique is a full connected subset of vertices of a graph; such a subset with $k$ vertices is often called a $k$-clique. In their model, two $k$-cliques are regarded as adjacent, if they share $k-1$ vertices. As classical percolation in Erd\H{o}s-R\'{e}nyi (ER) graphs, $k$-clique percolation considers the behavior of clusters formed by connected $k$-cliques for a different connection probability $p$ of edges. In network science, this approach has been proven successful in many applications for the analysis of network structure, such as the study of cancer-related proteins in protein interaction networks\cite{Jonsson15092006}, the analysis of stock correlations\cite{Heimo2007147}, and social networks\cite{palla2007quantifying}.

As a percolation model, clique percolation itself provides a set of very interesting problems. The simulation results in ref.\cite{PhysRevLett.94.160202} indicate that the fraction of vertices in the largest $k$-clique cluster makes a discontinuous percolation transition with increasing connection probability $p$; however, the fraction of $k$-cliques in the largest $k$-clique cluster demonstrates a continuous percolation transition. Bollob\'{a}s and Riordan give a rigorous mathematical result of the critical point for a more general percolation model, in which two $k$-cliques are regarded as adjacent if they share $l (=1,2,...,k-1)$ or more vertices\cite{RSA20270}. The Monte Carlo simulation indicates that for different $k$ and $l$, this general clique percolation model could demonstrate different critical phenomena\cite{Fan2014}. Moreover, the percolation of weighted and directed cliques is also studied as a community finding method in real-world networks\cite{Farkas2007,Palla2007}.

In this paper we will present a theoretical study of the general clique percolation in ER graphs and give a theoretical support and clarification of previous studies. The paper is organized as follows. In the next section some distribution characteristics of cliques in ER graphs will be obtained. In Sec. III, we will solve the clique percolation model. In Sec. IV, we examine the finite size scaling at the critical point. In the last section we summarize our findings.

\section{clique distribution}

For the study of the percolation in highly clustered networks, one usually needs to know the distribution characteristics of the clusters \cite{PhysRevE.80.036107,PhysRevE.82.066118,PhysRevE.87.062801}, so we begin our study with the clique distribution.A $k$-clique is composed by $k$ vertices and $\binom{k}{2}$ edges. For ER graphs, each possible edge occurs independently with probability $0<p<1$, so the expected number of $k$-cliques in an ER graph with $N$ vertices is
\begin{equation}
C=\binom{N}{k} p^{\binom{k}{2}}, \label{C}
\end{equation}
where $\binom{N}{k}$ gives the total number of cliques that can be formed in the graph. For $N\rightarrow \infty$, eq.(\ref{C}) can be rewritten as
\begin{equation}
C=\frac{N^k}{k!}p^{k(k-1)/2}. \label{CN}
\end{equation}
On average, the number of $k$-cliques containing a given vertex is
\begin{equation}
z_k=\frac{Ck}{N}=\frac{N^{k-1}}{(k-1)!}p^{k(k-1)/2}.
\end{equation}
For simplicity, we define $z_k$ as the average $k$-clique degree of vertices. Moreover, the $k$-clique degree distribution can be written as
\begin{equation}
g_k(i)= \binom{\binom{N-1}{k-1}}{i}\left[p^{\binom{k}{2}}\right]^i\left[1-p^{\binom{k}{2}}\right]^{\binom{N-1}{k-1}-i}, \label{gki}
\end{equation}
where $\binom{N-1}{k-1}$ gives the maximum number of $k$-cliques that could contain a given vertex. When $N\rightarrow \infty$, $g_k(i)$ takes the form
\begin{equation}
g_k(i)=\frac{z_k^i}{i!}e^{-z_k}. \label{gi}
\end{equation}
For $k=2$, $g_k(i)$ is the degree distribution of ER graphs and $z_2=Np$ is the average degree.

Following an edge, there is only one vertex. However, following an $i$-clique, we may find more than one $k$-clique or none ($i<k$). It is readily known that in ER graphs, a $k$-clique can be built from an $i$-clique with probability $p^{\binom{k}{2}-\binom{i}{2}}=p^{(k-i)(k+i-1)/2}$. Therefore, the probability distribution of the number of new $k$-cliques reached by following an $i$-clique of a $k$-clique can be written as
\begin{equation}
f_{k,i}(j)=\binom{\binom{N-k}{k-i}}{j}\left[p^{(k-i)(k+i-1)/2}\right]^j\left[1-p^{(k-i)(k+i-1)/2}\right]^{\binom{N-k}{k-i}-j}, \label{fkl}
\end{equation}
where $\binom{N-k}{k-i}$ gives the total number of choices to build a new $k$-clique from an $i$-clique of a $k$-clique. When $N\rightarrow \infty$, $f_{k,i}(j)$ also obeys the Poisson distribution
\begin{equation}
f_{k,i}(j)=\frac{y_{k,i}^j}{j!}e^{-y_{k,i}}. \label{fi}
\end{equation}
Here, $y_{k,i}=\sum_jf_{k,i}(j)j=N^{k-i}p^{(k-i)(k+i-1)/2}/(k-i)!$ is the average number of new $k$-cliques reached by following an $i$-clique of a $k$-clique.

\section{general formalism}

\begin{figure}
\scalebox{0.4}[0.4]{\includegraphics{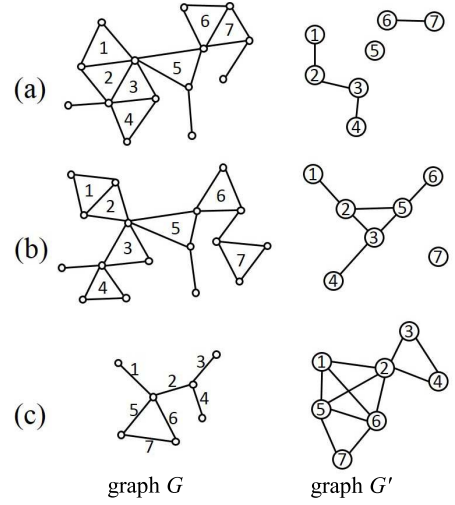}} \caption{Sketches of the mapping from clique percolation to nomal percolation. (a) $k=3$ and $l=2$. (b) $k=3$ and $l=1$. (c) $k=2$ and $l=1$. Vertices of graph $G'$ correspond to $k$-cliques of graph $G$ and two vertices of graph $G'$ are connected if the corresponding $k$-cliques of graph $G$ share $l$ or more vertices . Note that not all the vertices in graph $G'$ can be connected directly when edges are added in graph $G$, such as vertices $5$ and $7$ of graph $G'$ in (b). Therefore, graph $G'$ can not be entirely considered an ER graph, and the two graphs are not equivalent even for $k=2$ and $l=1$. Specially, when graph $G$ is a full connected graph, graph $G'$ will have a special topology.} \label{f1}
\end{figure}

Let $G$ be the ER graph we focus on. Then we consider a graph $G'$, whose vertices are the $k$-cliques of graph $G$. The vertices in graph $G'$ are connected if the corresponding $k$-cliques of graph $G$ share $l$ vertices or more (see Fig.\ref{f1}). Obviously, the clique percolation in graph $G$ corresponds to a normal percolation in graph $G'$.

An $i$-clique ($l \leq i <k$) of a $k$-clique in graph $G$ could map to more than one edge or none in graph $G'$, which satisfies the distribution (\ref{fi}). Therefore, the degree distribution of graph $G'$ is also a Poisson distribution with average degree $z'=\sum_{i=l}^{k-1}\binom{k}{i}y_{k,i}$. However, it is necessary to point out that graph $G'$ can not be entirely considered an ER graph (see Fig.\ref{f1}).

Furthermore, when the average degree $z'$ takes a finite size, the leading order of $z'$ must have size $y_{k,l} \sim \mathcal{O}(1)$. This yields $\sum_{i=l+1}^{k-1}\binom{k}{i}y_{k,i}$ has the order of $O(N^{\frac{k-1-l}{k-1+l}})$, which will vanish for large $N$. So when we consider the connection of graph $G'$, the edges mapped from $i$-cliques with $i>l$ can be ignored and the average degree of graph $G'$ can be simplified as $z'=\binom{k}{l}y_{k,l}$.

Next we calculate the critical point based on the branch process of graph $G'$. As we know, if a giant cluster exists in graph $G'$, the reproduction number of the branch process in such graph must be larger than one\cite{PhysRevE.64.026118,bollobas2001random}. That is, when we arrive at a vertex by following an edge, there must be excess edges leading to other vertices. Considering graph $G$, a $k$-clique has $\binom{k}{l}-1$ excess $l$-cliques and each $l$-clique could map to $y_{k,l}$ edges of graph $G'$ on average. Thus, the critical point satisfies
\begin{equation}
\left[\binom{k}{l}-1\right]y_{k,l}=1, \label{pc}
\end{equation}
that is
\begin{equation}
p_c = \zeta N^{-\frac{2}{k+l-1}}, \label{pc2}
\end{equation}
where $\zeta=\left[ \frac{(k-l)!}{\binom{k}{l}-1} \right]^{\frac{2}{(k-l)(k+l-1)}}$. This result is the same as the one found in ref.\cite{RSA20270}. When $l=k-1$, $p_c$ reduces to the critical point $[(k-1)N]^{-1/(k-1)}$ obtained in refs.\cite{Palla2007JSP,Rath2008}. For $k=2$ and $l=1$, $p_c=1/N$ is the critical point of the classical percolation in ER graphs. In addition, for $l<k-1$, if the edges mapped from $i$-cliques with $i>l$ are considered, a correction term $O(N^{-\frac{k-l+1}{k+l-1}})$ should be added at the critical point (\ref{pc2}).

The order parameter usually used in percolation theory is the fraction of vertices in the giant cluster; for clique percolation, it is the fraction of vertices in the giant clique cluster, labeled $\phi$. As pointed out in ref.\cite{PhysRevLett.94.160202}, there is another order parameter $\psi$ that can be used in clique percolation, which gives the fraction of cliques in the giant clique cluster. Note that $\phi$ can also be understood as the probability that a randomly chosen vertex belongs to the giant clique cluster and $\psi$ is the probability that a randomly chosen clique belongs to the giant clique cluster. To resolve the clique percolation, we need an auxiliary parameter $\varphi$, which expresses the probability that an $l$-clique of a $k$-clique can lead to the giant clique cluster. As the branch process discussed above, the giant cluster in graph $G'$ has tree-like structure near the critical point, so the connected cliques in graph $G$ must also have tree-like structure\cite{PhysRevLett.94.160202,Palla2007JSP,RSA20270}.

Following an $l$-clique of a $k$-clique, if we can not find the giant clique cluster, all the new $k$-cliques we find can not lead to the giant clique cluster. Thus, we can obtain a self-consistent equation for $\varphi$,
\begin{equation}
1-\varphi= \sum_{j=0}f_{k,l}(j)\left[(1-\varphi)^{\binom{k}{l}-1} \right]^j, \label{varphi}
\end{equation}
where $(1-\varphi)^{\binom{k}{l}-1}$ is the probability that all the other $l$-cliques of the $k$-clique reached by following the $l$-clique can not connect to the giant clique cluster. Substituting eq.(\ref{fi}) into eq.(\ref{varphi}), we have
\begin{eqnarray}
\varphi &=& 1- e^{y_{k,l}[(1-\varphi)^{\binom{k}{l}-1} -1]}  \nonumber  \\
&=& 1- e^{(p/p_c)^{\frac{(k-l)(k+l-1)}{2}}\left[ \binom{k}{l}-1\right]^{-1} \left[(1-\varphi)^{\binom{k}{l}-1} -1\right]}. \label{varphi2}
\end{eqnarray}
Solving this equation for the condition of $\varphi\rightarrow 0$, we will also find the critical point (\ref{pc2}).

Next we obtain the order parameters $\phi$ and $\psi$ by using $\varphi$. For a randomly chosen $k$-clique that does not belong to the giant clique cluster, all the $l$-cliques belonging to this $k$-clique can not lead to the giant clique cluster. Thus, $\psi$ satisfies
\begin{equation}
1-\psi= (1-\varphi)^{\binom{k}{l}}. \label{psi}
\end{equation}
Together with eq.(\ref{varphi2}), we find that the order parameter $\psi$ makes a continuous phase transition at the critical point (\ref{pc2}).

\begin{figure}
\scalebox{0.4}[0.4]{\includegraphics{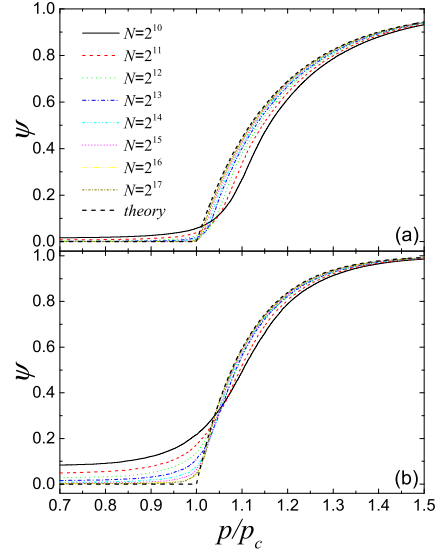}} \caption{(Color online) Simulation results of the order parameter $\psi$ for different graph sizes $N$. (a) $k=3$ and $l=2$. (b) $k=3$ and $l=1$. For both cases, the order parameter $\psi$ demonstrates a continuous phase transition. The theory prediction is obtained from eqs.(\ref{varphi2}) and (\ref{psi}).} \label{f2}
\end{figure}

The simulation results of the order parameter $\psi$ for $k=3$ are shown in Fig.\ref{f2}, which are in good agreement with the theory prediction of eqs.(\ref{varphi2}) and (\ref{psi}). One can find that the behavior of the order parameter $\psi$ is much the same as that of the classical percolation. This is because that the order parameter $\psi$ just describes a normal percolation in the mapped graph $G'$.

We now turn our attention to the order parameter $\phi$ that is usually used in percolation theory. For a randomly chosen vertex that does not belong to the giant clique cluster, all the $k$-clique degrees of the vertex can not lead to the giant clique cluster. Thus, $\phi$ satisfies
\begin{equation}
1-\phi= \sum_{j=0} g_k(j) \left[(1-\varphi)^{ \binom{k-1}{l}}\right]^j. \label{phi}
\end{equation}
Note that the probability in the square brackets of eq.(\ref{phi}) should express the excess connection of the $k$-clique degree of the vertex, so the $l$-cliques containing the vertex are not counted. Substituting eq.(\ref{gi}) into eq.(\ref{phi}), we have
\begin{equation}
\phi = 1-e^{z_k[(1-\varphi)^{\binom{k-1}{l}}-1]}.   \label{phi2}
\end{equation}
Specifically, for $k=2$ and  $l=1$, eqs.(\ref{varphi2}) and (\ref{phi2}) are just the equations for the classical percolation in ER graph\cite{PhysRevE.64.026118}.

For the discussion of the case $N\rightarrow \infty$ near the critical point, we rewrite eq.(\ref{phi2}) as
\begin{equation}
\phi = 1-e^{  (p/p_c)^{\binom{k}{2} } \xi w(N) \left[ (1-\varphi)^{\binom{k-1}{l}}-1 \right]},   \label{phi3}
\end{equation}
where
\begin{eqnarray}
\xi  &=& \frac{1}{(k-1)!} \left[ \frac{(k-l)!}{\binom{k}{l}-1} \right]^{\frac{k(k-1)}{(k-l)(k+l-1)}}, \\
w(N) &=& N^{\frac{(k-1)(l-1)}{k+l-1}}.
\end{eqnarray}
For $l=1$, $w(N)=1$ and then $\phi_c \sim \varphi_c$, which indicates that $\phi$ makes a continuous phase transition as $\varphi$. For $l>1$, $\phi_c \sim w(N)\varphi_c$. This means that the critical behavior of $\phi$ depends on the relative sizes of $w(N)$ and $\varphi_c$.

\begin{figure}
\scalebox{0.4}[0.4]{\includegraphics{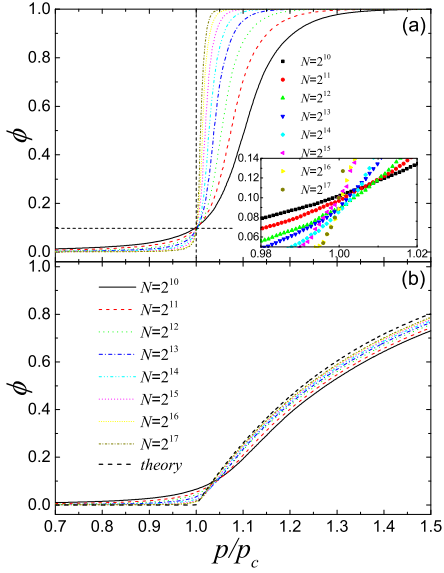}} \caption{(Color online)Simulation results of the order parameter $\phi$ for different graph sizes $N$. (a) $k=3$ and $l=2$. The phase transition of the order parameter $\phi$ becomes discontinuous in the thermodynamic limit. The inset shows the details close to the percolation threshold. (b) $k=3$ and $l=1$. The order parameter $\phi$ demonstrates a continuous phase transition. } \label{f3}
\end{figure}

As shown in Fig.\ref{f3}, the phase transition of $\phi$ for $l=1$ demonstrates a continuous phase transition, which is agreement with our theoretical prediction well when $N\rightarrow \infty$. For $l>1$, the phase transition of the order parameter $\phi$ tends to be a step function for $N\rightarrow \infty$ with a nearly constant $\phi_c$ for different graph sizes $N$. An explanation will be given when we get the finite size scaling of $\varphi_c$ in the next section. Otherwise, it should be noted that even though the model reduces to the classical percolation $(k=2, l=1)$, graphs $G$ and $G'$ are also different (see Fig.\ref{f1} (c)), so the order parameters $\psi$ and $\phi$ are always inequivalent.

\section{at critical point}

It is known that, in ER graph, when $p<p_c$ almost surely each cluster has size $logN$, at $p=p_c$, the largest cluster has size of order $N^{2/3}$, and at $p>p_c$, there exists a single giant cluster of size of order $N$ and all other clusters have size of order $logN$\cite{bollobas2001random}. As discussed above, when a clique percolation occurs in graph $G$, there is a corresponding normal percolation in the mapped graph $G'$. Thus, the number of cliques in the giant clique cluster could also satisfy \cite{PhysRevLett.94.160202}
\begin{equation}
C^*\sim\left\{
      \begin{array}{ll}
        logC, & p<p_c, \\
        C^{2/3}, & p= p_c, \\
        C, & p>p_c.
      \end{array}
    \right.  \label{cs}
\end{equation}
Using eqs.(\ref{C}) and (\ref{pc2}), we can find the total number of cliques at $p=p_c$,
\begin{equation}
C_c \sim N^{\frac{kl}{k+l-1}} .     \label{cc}
\end{equation}
Then, the giant clique cluster at the critical point $C^*_c$ scales as $N^{2kl/3(k+l-1)}$. Note that the giant clique cluster has tree-like structure at the critical point, which means the number of cliques in the giant clique cluster $C^*_c$ can not grow faster than the number of vertices. This yields
\begin{equation}
C^*_c \sim \left\{
              \begin{array}{ll}
                N^{2kl/3(k+l-1)} , & \frac{2kl}{3(k+l-1)}<1 \\
                N, & \frac{2kl}{3(k+l-1)} \geq1.
              \end{array}
            \right.
\end{equation}
Together with eq.(\ref{cc}), we have
\begin{equation}
\psi_c =\frac{C^*_c}{C_c}\sim \left\{
              \begin{array}{ll}
                N^{-kl/3(k+l-1)} , & \frac{2kl}{3(k+l-1)}<1, \\
                N^{-(k-1)(l-1)/(k+l-1)}=\frac{1}{w(N)}, & \frac{2kl}{3(k+l-1)}\geq1.
              \end{array}
            \right. \label{psic}
\end{equation}

\begin{figure}
\scalebox{0.4}[0.4]{\includegraphics{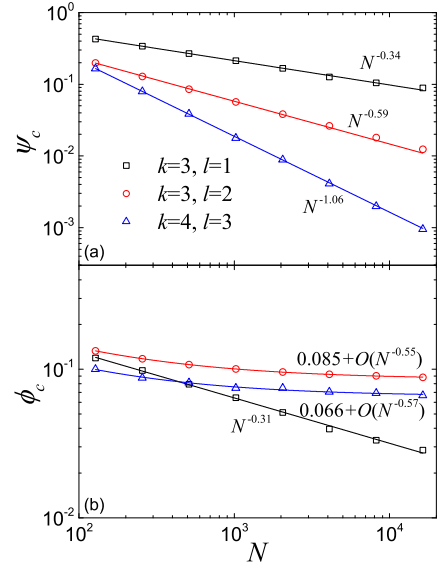}} \caption{(Color online) Simulation results of the order parameters at the critical point. Here $\psi_c$ and $\phi_c$ scale as a power of graph size $N$. The corresponding theoretical results are as follows: (a) $\psi_c(k=3,l=1)\sim N^{-1/3}$, $\psi_c(k=3,l=2)\sim N^{-1/2}$, and $\psi_c(k=4,l=3)\sim N^{-1}$ and (b) $\phi_c(k=3,l=1)\sim N^{-1/3}$. For $l>1$, $\phi_c$ behaves as $\phi_c=\phi_{c0}+O(N^{-\epsilon})$ for finite size systems.} \label{f4}
\end{figure}

From eq.(\ref{psi}), we find that the order parameter $\psi_c \sim \varphi_c$, that is, $\varphi_c$ also satisfies eq. (\ref{psic}). Then, replacing $\varphi$ by the order of $N$ in eq.(\ref{phi3}), we have
\begin{equation}
\phi =  \left\{
              \begin{array}{ll}
                0, & p<p_c  \\
                1-e^{ \kappa }, & p= p_c \\
                1, & p>p_c.
              \end{array}
            \right. \label{phic}
\end{equation}
Here, $\kappa$ is a constant determined by $k$ and $l$. This means that the order parameter $\phi$ makes a special phase transition that when $p<p_c$, $\phi=0$; when $p>p_c$, $\phi=1$; and when $p=p_c$, the order parameter is a constant $0<\phi_c<1$. Furthermore, when $l=1$, $\phi_c \sim \varphi_c$, i.e., $\phi_c$ scales as the same power of $N$ with $\psi_c$.

In Fig.\ref{f4}, we give the finite size scaling at the critical point. In accord with our analysis, the order parameters $\phi_c$ for $l=1$ and $\psi_c$ are in agreement with eq.(\ref{psic}) and the order parameter $\phi_c$ at the critical point for $l>1$ is a constant for $N\rightarrow \infty$. For finite sizes, $\phi_c$ for $l>1$ behaves as $\phi_c=\phi_{c0}+\mathcal{O}(N^{-\epsilon})$ (see Fig.\ref{f4} (b)).

In addition, for large $k$, the edges from the high-order term of $y_{k,i}$ ($i>l$) will have a role in the percolation process of a finite size system, so the finite size scaling discussed above gives only approximate results for the case of large $k$ and $l<k-1$. However, for $l=k-1$, all the results discussed above are accurate (see the case of $k=4$ and $l=3$ in Fig.\ref{f4}). Furthermore, another possible reason for the deviation between the theoretical results and the simulation results in Fig.\ref{f4} is that graph $G'$ can not be identical to an ER graph.

\begin{figure}
\scalebox{0.3}[0.3]{\includegraphics{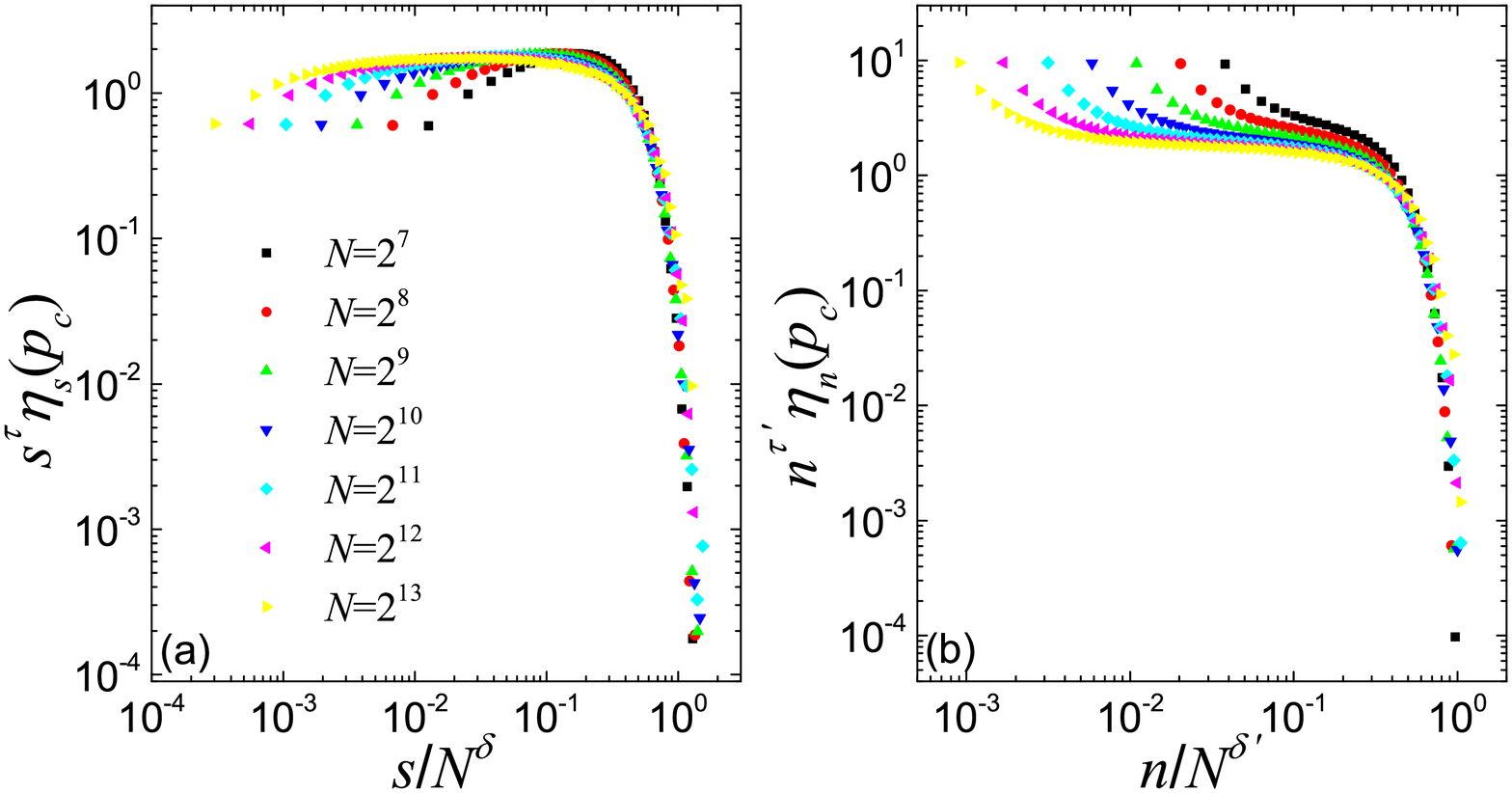}} \caption{(Color online) Simulation results of the clique cluster number distribution at the critical point, where a clique cluster size is measured by the number of (a) cliques and (b) vertices in the cluster. For these plots, $\tau=\tau'=5/2$ and $\delta=\delta'=0.9$ are used.} \label{f5}
\end{figure}

To clarify the discontinuous percolation transition of $\phi$ for $l>1$, the clique cluster number distributions at the critical point are shown in Fig.\ref{f5}. The scaling ansatz are set as $\eta_s(p_c)\sim s^{-\tau} f(s/N^\delta)$ and $\eta_n(p_c) \sim n^{-\tau'} f(n/N^{\delta'})$. Here, $s$ means the number of cliques in a clique cluster, and $n$ the number of vertices. From Fig.\ref{f5}, we find that the two distributions give the same critical exponents $\tau=\tau'=5/2$ and $\delta=\delta'\simeq 0.9$. This indicates that there is no essential distinction between the two percolation transitions measured by $\psi$ and $\phi$. The different behaviors of the order parameters $\psi$ and $\phi$ are mainly caused by the quantitative relation between the total number of cliques and the total number of vertices.

\section{conclusion}

In this paper, we have established three mean field equations for clique percolation in random graph. Using these equations, we not only covered the critical points found in previous studies, but also obtained the corresponding order parameter in the thermodynamic limit. Based on this, we prove theoretically that the order parameter $\phi$ for $l>1$ makes a step-function-like phase transition in the thermodynamic limit and a continuous phase transition for $l=1$. More interesting, our analysis showed that in the thermodynamic limit, the order parameter $\phi$ is neither $0$ nor $1$, but a constant depending on $k$ and $l$. Through simulation we found that the order parameter $\phi$ behaves as $\phi_c=\phi_{c0}+O(N^{-\epsilon})$ at the critical point, which supports our theoretical study. In addition, our analysis also proved that the order parameter $\psi$ always makes a continuous phase transition as the classical percolation. Through analysis of the clique cluster number distribution, we found that there is no essential distinction between the two processed measured by $\psi$ and $\phi$ and the different behaviors are mainly caused by the quantitative relation between the total numbers of cliques and vertices.

Although only ER graphs were studied in this paper, a similar discussion can be had for other networks if we know the distributions $f_{k,i}(j)$ and $g_k(i)$. These results give a theoretical clarification for the clique percolation, which will enhance the understanding of the emergence of community structures in complex networks and the clique percolation itself.

\begin{acknowledgments}

The research of M.L. was supported by the Fundamental Research Funds for the Central Universities and the National Natural Science Foundation of China under Grant No. 61503355. The research of Y.D. was supported by the National Natural Science Foundation of China under Grant No. 11275185 and the Chinese Academy of Sciences. Y.D. also acknowledges support from the Specialized Research Fund for the Doctoral Program of Higher Education under Grant No. 20113402110040 and the Fundamental Research Funds for the Central Universities No. 2030020028. The research of B.-H.W. was supported by the National Natural Science Foundation of China under Grant No. 11275186 and the Open Funding Programme of Joint Laboratory of Flight Vehicle Ocean-based Measurement and Control under Grant No. FOM2014OF001.

\end{acknowledgments}

\bibliography{ref}

\end{document}